\documentclass{aa}
\usepackage{graphicx}
\usepackage{times}
\usepackage{astron}

\begin{document}

\thesaurus{3   11.05.2;  
               11.09.5;  
               11.11.1;  
               11.19.6;  
           } 

\title{The faint end of the Tully-Fisher relation}

\author{J.M. Stil\inst{1}
\and F.P. Israel\inst{1}}
\institute{Leiden Observatory, PO Box 9513, 2300 RA Leiden, The Netherlands}

\offprints{J.M. Stil}
\mail{stil@strw.leidenuniv.nl}

\date{Received 11 September 1998; accepted 0000}
\maketitle

\begin{abstract}

We have studied the relation between HI linewidth and optical luminosity 
in two samples of dwarf galaxies, with special attention to
the accuracy of the inclination correction. 
The first sample consists of rotationally supported galaxies, and has
DDO~154 as a prototype. They have HI linewidths significantly larger than
predicted by an extrapolation of the Tully-Fisher relation.
The deviation from the extrapolated Tully-Fisher relation correlates with 
the distance-independent mass-to-lumi\-no\-sity ratio $M_{\rm HI}/L_{\rm B}$. 
The baryonic correction for these gas rich galaxies is too small to 
explain the deviation from the Tully-Fisher relation, unless
a small mass-to-light ratio $(M/L_{\rm B})_*$ $\le 0.25$ for the stellar 
population is assumed.
The second sample contains dwarf galaxies with comparable rotational and 
random velocities. This sample, having Leo A as a prototype, is consistent 
with an extrapolation of the classical Tully-Fisher relation to lower 
luminosities. These dwarf galaxies are not particularly rich in gas, but
their mass is dominated by baryonic matter within their HI radius. 
We suggest that any evolution of the deviating galaxies towards the Tully-Fisher 
relation will combine a decrease in linewidth by radial transport of HI 
with an increase in luminosity due to a larger accumulated stellar mass.

\end{abstract}

\section{Introduction}

The luminosity $L$ and the rotational velocity $v$ of a galaxy
are empirically related by the so-called Tully-Fisher relation (Tully \&
Fisher 1977). Simple arguments lead to the general form $L \sim v^4$, 
although the value of the exponent depends on the filter in which $L$ 
is measured as well as the manner of determining
the rotational velocity $v$ (Verheijen 1997). Although it is often 
assumed that the relation is strictly linear in a logarithmic sense, 
there is no reason why this should apply to all galaxies.

A useful reference frame is provided by the Tully-Fisher relation from
Kraan-Korteweg et al. (1988), fitted to a sample of 13 nearby galaxies
in the luminosity range $ -16.4 > M_{\rm B} > -21.7$:
$$
M_{\rm B}=-6.69~\ {\rm log\Delta} v_{21} -2.77\ \ \ (\pm 0.1) \eqno (1)
$$
where $\Delta v_{21}$ is the {\it inclination-corrected} width of
the 21 cm HI line profile at 20\% of the peak value; we reserve the
symbol $W_{20}$ for the {\it observed} linewidth at 20\% of the peak value. 
The calibration of this relation corresponds to a Hubble constant 
$H_0=56.6\ \rm km\ s^{-1}\ Mpc^{-1}$. Kraan-Korteweg et al. (1988) have
compared this relation to the Tully-Fisher relation defined by different 
types of galaxies in the Virgo cluster. The dwarf galaxies discussed here are
considerably fainter than any of the galaxies considered by Kraan-Korteweg et
al. (1988).  We must therefore extrapolate the Tully-Fisher relation to lower 
luminosities. As an example, a galaxy with linewidth $ \Delta v_{21}=50\ 
\rm km\ s^{-1}$ is thus predicted to have a luminosity $M_{\rm B}\approx -14$.

In the literature, observations exist of half a dozen dwarf galaxies
fainter than $M_{\rm B} = -14.5$, with rotational velocities up to
$50\rm \ km\ s^{-1}$ (i.e. $\Delta v_{21}\approx 100$) and with
reliable distances (DDO~154: Carignan \& Beaulieu 1989; DDO~161,
UGCA~442, ESO381-G20: C\^ot\'e et al. 1997; DDO~87: Stil \& Israel, in
preparation).  Until now, only two objects were known to deviate
by a large amount from the extrapolated Tully-Fisher relation: DDO~154
(Carignan \& Beaulieu 1989) and the more luminous starburst dwarf
NGC~2915 (Meurer et al. 1996). Both have luminosities lower than
expected from their linewidth and the the T-F relation. Although Zwaan
et al. (1995) found no difference between the T-F relations pertaining
to more luminous low-surface brightness galaxies and to high-surface
brightness galaxies, Matthews et al. (1997) concluded that extreme
late-type galaxies also are systematically fainter than inferred from
the T-F relation and their linewidth. Furthermore,
Matthews et al. (1997) found the baryonic correction, i.e. a
correction to the luminosity to account for the large gas fraction in these
galaxies, too small to explain the observed deviation.

Here we present two samples
of dwarf galaxies, all observed in detail with a radio interferometer,
extending the low-luminosity limit to $M_{\rm B}$ = -13 and -10
respectively. The first, containing galaxies clearly dominated by
rotation, deviates significantly from the extrapolated T-F
relation. In contrast, the second sample of dwarf galaxies with
considerably smaller rotational velocities does not, at least over the
same luminosity range. The physical differences between the two
samples are discussed.

\section{Sample selection}

\subsection{General considerations}

We have studied two samples of dwarf galaxies which are distinguished by
their kinematic properties:
(1). dwarf galaxies with a regular velocity field, kinematically 
well-described by tilted ring models, and (2). dwarf galaxies rotating 
so slowly that the random motions of the HI are dynamically important. 
As the dynamical mass is roughly proportional to $R_{\rm HI} v^2$, 
where $R_{\rm HI}$ is the radius of the HI distribution and $v$ is
the rotation velocity at that radius,
these samples are also expected to be different in dynamical mass. 
Sample galaxy distances were taken from published stellar 
photometry or published group membership. Hubble
expansion and a correction for Virgocentric infall were adopted for
a small fraction of the sample. 

Galaxy linewidths $W_{20}$ were taken from Bottinelli et~al. (1990), 
insuring that the corrections for instrumental resolution and primary beam 
dilution were performed in the same way for all objects.

\subsection{Rotation curve (RC) sample}

The first sample consists of galaxies with an accurately measured rotation 
curve. The selection criteria for this rotation-curve sample were:
(1). the rotation curve can be determined by a tilted-ring analysis; (2).
the rising part of the rotation curve is well resolved; (3). the inclination
can be included in the fit as a free parameter.

The use of a tilted ring analysis does not allow inclusion of objects with an 
inclination $i<50^{\circ}$ and objects with a solid-body rotation curve, 
as their inclination cannot be determined from the tilted ring fit. 
The advantage of obtaining kinematic inclinations is that it avoids 
assumptions about the poorly known intrinsic axial ratios of these objects.
Nevertheless, we have included in Table~\ref{sample1} four dwarf galaxies 
with considerable rotational velocity for which the inclination could not be 
determined from a tilted ring fit. For these inclinations were estimated from 
the ellipticity of the HI isophotes. One of these galaxies, DDO~47 is a
particularly good example of a galaxy with a solid-body rotation curve.  

Although we limit our analysis to the dwarf galaxies in this sample, i.e.
galaxies with $M_{\rm B}>-16$, we have also included in Table~\ref{sample1}
a number of more luminous galaxies for comparison purposes.

\begin{table*}
\caption{Rotation curve sample with fast rotators}
\small
\tabcolsep=1.5mm
\begin{tabular}{|  l  | l | c | l | r | c | c | c | l |  c |  c |  c | c | l | l |} 
\hline 
  Name   &  dist &  $\rm M_B$& B-V & h & $\rm R_{max}$ & $\rm V_{max}$ &$<\sigma>$ & Type & $i$ & $\rm \Delta v_{21}$ &$\rm M_{HI}/L_B$ &$\rm {\rm log}(M_{VT}/\beta)$ &ref \\ 
         &  Mpc  &       &      &kpc&  kpc    &$\rm km\ s^{-1}$&$\rm km\ s^{-1}$& & $^\circ$  & $\rm km\ s^{-1}$&  &   &  \\
 \ \ \ \ \ \ \ [1]  &\ \ [2] &  [3]  &  [4] &   [5]     & [6]  & [7] & [8] & [9] &[10] & [11] & [12] & [13] & [14] \\ 
\hline
DDO  87   &  3.4 &  -12.8 &0.26&     & 2.0 & 34.4 & 6.0&Im    &$63\pm2$& 82&2.6  & 8.77& a  \\ 
DDO 154   &  4   &  -13.8 &0.37& 0.50& 7.6 & 43.1 &    &IBm(s)&$64\pm3$&109&5.5  & 9.57& c *  \\ 
UGCA 442  & 2.5  &  -13.8 &0.29& 0.43& 3.6 & 54.4 & 8.4&SBm(s)(sp)&\ \ $70\pm10$&120&1.6& 9.43& m * \\
ESO381-G20& 3.5  &  -13.9 &0.3 & 0.62& 3.7 & 62.8 & 8.7&IBm(s) &$58\pm4$&121&1.7 & 9.56& m * \\
DDO  83   &  9   &  -14.5 &0.01& 0.65& 3.3 & 51.0 &10.0&IBm(s)&$53\pm2$&150&2.4& 9.34& a    \\ 
DDO 170   &  12  &  -14.5 &0.4 & 1.30&11.7 & 62.2 &    &Im    &$70\pm3$&82 &4.8  &10.05& d * \\ 
DDO 161   & 3.5  &  -14.9 &0.29& 0.70& 6.2 & 64.4 & 9.3&IBm(s) &$68\pm5$&147&2.3 & 9.80& m * \\
UGC 7906  & 16   &  -15.0 &0.43& 1.24& 4.6 & 55   &    &Im     &$50\pm5$&151&1.6 & 9.54& k * \\
DDO 168   &  3.5 &  -15.2 &0.22& 0.92& 3.4 & 48.7 &10.6&IBm   &$61\pm1$& 90&1.1  & 9.33& a,b *\\ 
IC 3522   & 16   &  -15.6 &0.53& 1.62& 6.9 & 57.0 &    &IBm  &$70\pm5$&128&2.2 & 9.75 & k * \\
NGC 1560  & 3.0  &  -15.9 &0.57& 1.3 & 8.3 & 76.6 &    &SAd(sp)&$80\pm3$&159&2.5 &10.07& b *\\ 
NGC 2915  & 5.3  &  -15.9 &0.39& 0.66&14.1 & 87.0 &12.0&I0    &$57\pm2 $&199&2.9&10.42& h * \\ 
DDO  48   &  16  &  -16.4 &0.59& 1.78& 8.0 & 76.6 &10.1&Sm    &$80\pm4$&168&1.3   &10.06& a \\ 
NGC 3109  &  1.7 &  -16.8 &    & 1.55& 8.2 & 65.7 &    &SBm(s) &$70\pm5$&128&1.7   & 9.94& n *\\ 
IC 3365   & 16   &  -17.0 &0.52& 1.58& 4.6 & 58   &    &Im   &$70\pm5$&156&0.3 & 9.58& k * \\
UGC 4278  & 6.4  &  -17.3 &0.40&     & 4.7 & 87.6 &    &SBd(sp)  &$90    $&201&0.35& 9.94& a  \\
NGC 2976  & 3.4  &  -17.4 &0.58& 1.07& 2.0 & 71.5 &11.8&SAcP  &$66\pm2$ &183&0.11& 9.41& a \\ 
NGC 5585  & 6.2  &  -17.5 &0.47& 1.4 & 9.7 & 89.4 &    &SABd(s)&$52\pm2$&209&1.1 &10.26& i *  \\ 
NGC 300   & 1.8  &  -17.9 &0.58& 2.06&10.6 & 93.4 &    &SAd(s)&$52\pm8$&190&0.33 &10.34& f *  \\ 
NGC 247   & 2.5  &  -18.0 &0.54& 2.93& 9.8 &107.8 &    &SABd(s)&$74\pm3$&229&0.34&10.43& e *   \\ 
DDO 105   &  16  &  -18.1 &0.53&     &15.1 & 81.8 &    &IBm(s)&$64\pm2$&187&0.8  &10.38& b    \\ 
M 33      & 0.72 &  -18.6 &0.47& 1.7 & 5.4 &106.8 &    &Scd   & $54$     &251&0.24&10.16& l  \\
Dwingeloo 1 & 3  &  -19.0 &    &     & 6.4 & 109  &    &SBb    &$51\pm2$&262&0.073&10.27& j \\
NGC 2403  & 3.0  &  -19.3 &0.39& 2.05&19.5 & 134  &    &SABcd(s)&$60\pm2$&302&0.39 &10.91& g *  \\ 
NGC 3198  & 9.4  &  -19.4 &0.43& 2.60&10.9 & 153  &    &SBc(rs)&$74\pm3$&332&0.56 &10.78& g  * \\ 
NGC 2903  & 6.4  &  -20.0 &0.55& 2.02&14.9 & 188  &    &SABbc(rs)&$63\pm3$&429&0.16&11.09& g * \\ 
NGC 801   & 79   &  -21.7 &0.61& 12  & 59  & 216  &    &Sc    &$75\pm1$&468&0.30 &11.8& b *  \\ 
\hline
DDO 47    & 2    & -13.4  &0.35& 0.5 & 1.6 & 67.0 & 8.7&IBm(s) & 30 (HI)&158&1.8 & 9.24& a \\ 
DDO 52    & 5.3  & -13.8  &0.53& 0.6 & 3.0 & 52.9 & 7.4&Im: & 55 (HI)&123&1.1 & 9.31& a \\ 
DDO 68    & 6.1  & -14.3  &0.23& 0.5 & 3.5 & 54.7 &12.6&Im pec & 57 (HI)&116&2.9 & 9.45& a \\ 
DDO 64    & 6.1  & -14.7  &0.15& 1.2 & 2.7 & 41.8 &10.9&Im & 80 (HI)&110&1.2 & 9.11& a \\ 
\hline
\end{tabular}
Column definitions :
[1] Object name; galaxies are ordered by luminosity; objects below the horizontal line have no 
kinematic estimate of the inclination;
[2] distance in Mpc;
[3] absolute magnitude in B;
[4] B-V colour;
[5] exponential disc scale length (from the rotation curve reference);
[6] radius of the last measured point of the rotation curve in kpc;
[7] velocity of the last measured point of the rotation curve in kpc;
[8] mean HI velocity dispersion;
[9] galaxy type from RC3;
[10] inclination from the tilted ring fit in degrees;
[11] Full width of the 21 cm line profile at 20 \% of the peak 
divided by $\sin(i)$, in $\rm km\ s^{-1}$. Linewidths were taken from \cite{bottinelli}; 
[12] ratio of HI mass to B luminosity in $M_\odot/L_{\rm B \odot}$;
[13] ${\rm log}(\beta^{-1}(M_{\rm VT}/10^8 M_\odot))$ from equation (3);
[14] references to rotation curve data : a Stil \& Israel (in preparation); b Broeils (1992); 
c Carignan \& Beaulieu (1989); d Lake et al. (1990); e Carignan \& Puche (1990); f Puche et al. (1990); 
g Begeman (1987); h Meurer et al. (1996); i C\^ot\'e et al. (1991); j Burton et al. (1996); 
k Skillman et al. (1987); l Newton (1980); m C\^ot\'e et al. (1997); n Jobin \& Carignan (1990).
Detailed mass models exist in the literature for the galaxies indicated with *
\normalsize 
\label{sample1}
\end{table*}

\subsection{Slow-rotation (SR) sample}
\label{slowrot-sample}

The second sample is based on our aperture synthesis radio observations
of ``classic'' dwarf galaxies which exhibit little or no rotation,
very different from the rotation-curve (RC) sample discussed above.
The selection criterion for this sample is low observed maximum 
{\it projected} rotation velocity $V_{\rm max}~\ \sin(i) < 25\ \rm km\ s^{-1}$.
Since the faint dwarf galaxies studied by Lo et al. (1993) also obey this 
criterion, they are included in the sample.

The selection on {\it projected} rotation velocity neglects the effects of
the uncertain inclination of these objects. The sample may thus be contaminated
by some fast-rotating face-on galaxies, but we are confident that we can
identify these by the large inclination correction implied by their axial
ratios. The resulting slow rotators are listed in 
Table~\ref{slow-rot}. A 
lower limit of $30^{\circ}$ to the inclination is adopted. A justification
for this limit is given in Section~\ref{model-sec}.

We did not constrain the luminosity of the galaxies in this sample. The 
luminosities of the slow rotators, which may be as low as $M_{\rm B} = -10$, 
in fact overlap with the RC sample in the range $-12 > M_{\rm B} > -16$. The 
three faintest slow rotators are objects from the sample of Lo et al. (1993).

\begin{table*}
\caption{Slow-rotation sample}
\small
\tabcolsep=1.5mm
\begin{tabular}{|  l  | l |  r | r | l | c  c | c | c | c | c | c | l | l |} 
\hline 
  Name    &  dist  &  $\rm M_B$ &B-V & Type &   $i$   & src &$\rm W_{20}$  & $\rm R_{max}$ & $\rm V_{max}\sin(i)$ & $\rm M_{HI}/L_B$ & $\rm {\rm log}(M_{VT}/\beta)$ & ref  \\ 
          &  Mpc   &            &    &      &$^\circ$&     &$\rm km\ s^{-1}$ & kpc &$\rm km\ s^{-1}$ &  &  & \\
 \  [1]   &   [2]  &  [3]       &[4] &[5] & [6]  & [7] & [8] &[9] &[10] &[11]& [12] & [13] \\ 
\hline
LGS 3      & 0.76 &  -9.2 & 0.7 &I?       & 34  & opt   & 36 &0.5& $<2$      &0.28&6.73& b\\
DDO 210    & 1.0  &  -9.9 & 0.08&IAm      & 45  & opt   & 31 &0.3& $<5$      &2.21&6.84& b\\
Sag DIG    & 1.1  & -10.6 & 0.4\ &IBm(s)   & 40  & opt   & 31 &0.8& $<2$      &3.10&7.00& b\\
M81 dw A   & 4.6  & -12.4 & 0.4\ &I?       & 37  & opt   & 33 &1.0& $2.5\pm 1$      &0.81&7.67&g \\
UGC 4483   & 3.25 & -13.0 &     &Im:      & 30  & opt   & 56 &1.0& $10\pm 2$ &1.1 &8.09& b\\
DDO 216    & 1    & -13.1 & 0.62&Im       & 58  & opt/HI& 35 &0.6& $7\pm5$   &0.15&7.32& a,b \\ 
DDO 69     & 1.5  & -13.5 & 0.31&IBm      & 59  & opt/HI& 46 &1.1& $< 2 $    &0.96&7.30& b,f \\
DDO 75     & 1.7  & -13.8 & 0.37&IBm(s)   & 36  & opt/HI& 62 &1.2& $20 \pm 5$&1.18&8.57& d\\
DDO 187    & 4.0  & -13.8 & 0.38&Im       & 37  & opt   & 49 &1.1& $<5$      &1.02&7.79& b\\
IC 1613    & 0.725& -14.4 & 0.67&IBm(s)   & 38  & opt/HI& 35 &2.6& $12 \pm 2$&0.70&8.50& e\\
DDO 22     & 9.9  & -14.9 & 0.09&Im:      & 90  & HI    & 67 &2.9& $19\pm3$  &1.04&8.68& a \\ 
DDO 63     & 3.4  & -15.0 &-0.09&IABm(s)  & 42  & HI    & 44 &2.5& $8.6\pm3$ &0.73&8.34& a,c \\ 
Mkn 178    & 5.2  & -15.0 & 0.35&Im       & 56  & opt   & 47 &1.1& $\sim 6$ &0.12&7.65& a \\
DDO 125    & 4.5  & -15.6 & 0.40&Im       & 50  & opt/HI& 46 &2.6& $11\pm3$  &0.41&8.35& a,c \\ 
DDO 165    & 4.6  & -15.8 & 0.23&Im       & 38  & opt/HI& 62 &1.9& $26\pm7$  &0.33&9.00& a \\ 
\hline
\end{tabular}

Column definitions:
[1] Object name; DDO 63 = Holmberg I; DDO 69 = Leo A; 
DDO 75 = Sex A; IC 1613 = DDO 8; 
[2] distance in Mpc;
[3] absolute magnitude in the B band;
[4] B-V colour;
[5] galaxy type from RC3;
[6] inclination in degrees;
[7] source of the inclination : opt = optical axial ratio, HI = HI axial ratio;
assumed intrinsic axial ratio is $q_0=0.15$ as explained in the text;
[8] width of 21 cm line profile at 20 \% of the peak in $\rm km\ s^{-1}$, without inclination correction;
[9] measured maximum rotation velocity without inclination correction;
[10] ratio of HI mass to B luminosity in $M_\odot/L_{\rm B \odot}$;
[11] ${\rm log}(\beta^{-1}(M_{\rm VT}/10^8M_\odot))$ from equation (3);
[12] references : a Stil \& Israel (in preparation) ; b Lo et al.(1993); c Tully et al. (1978); 
d Skillman et al. (1988); e Lake \& Skillman (1989); 
f Young \& Lo (1996); g Sargent et al. (1983)
\normalsize
\label{slow-rot}
\end{table*}

\subsection{Comments on structure}

Some general remarks about the HI and optical structure of the dwarfs
in the two samples can be made. 

The dwarf galaxies in the RC sample usually have a large HI disc 
with modest HI column densities surrounding the optical galaxy. A
representative example is DDO~154 (Carignan \& Beaulieu 1989). 

The HI distribution of the slow rotators DDO~63, DDO~69, DDO~125, and
DDO~165 is characteristized by a small number of regions with higher
HI column density, usually in a ring or a ridge, around the edge of
the optical galaxy.  A smaller HI column density is observed near the
centre. DDO~63, DDO~69 and DDO~125 appear to have an extended HI halo
with a low HI column density. DDO~69 (Young \& Lo 1996) is a
representative example. DDO~22 is suggested as an edge-on slow
rotator because of its elongated HI distribution and its similar
appearance on the POSS plates.

\section{Correction for inclination}
\label{model-sec}

The actual HI linewidth $\Delta v_{21}$ of a galaxy is a complex
combination of the rotation curve, the gas distribution and the
relative importance of an isotropic component. The observed linewidth
is smaller as a result of projection onto the plane of the sky.

These factors are of minor importance in large spiral galaxies where
generally a symmetric HI disc extends far beyond the radius where the rotation 
curve reaches its high, constant value. In contrast, the gas
distribution and HI velocity dispersion are important in dwarf
galaxies which have a rising rotation curve and a low overall rotation
velocity. Effects of the HI distribution on HI linewidths were
studied before by Shostak (1977).

The relation between the observed linewidth $W_{20}$ and the intrinsic
linewidth $\Delta v_{21}$ is straightforward if the rotational
velocity is much larger than the isotropic random motions, so that
$W_{20} \approx \Delta v_{21}~\sin(i)$. More in general, however, a
correction for the contribution of random motions cannot be
avoided. This is often done in a statistical way described by
Tully \& Fouqu\'e (1985).  As this correction works in the direction
of the effect described in Section~\ref{TF-sec}, and therefore might
introduce spurious systematic differences between the two samples, we
have chosen not to apply it. Instead, we will model the magnitude of
the effects due to gas distribution, random motions and inclination on
the observed widths of the line profiles. In order to do so, we have
constructed a set of Monte Carlo model observations of dwarf galaxies
with the GIPSY program GALMOD (originally written by T.S. van Albada).
A model observation is based on a rotation curve, an axially symmetric
HI distribution and the velocity dispersion of the HI gas.  An HI
layer with finite thickness can be modeled. We adopted an exponential
vertical profile with a scale height above the plane of 10\% of the HI
radius. Changes in this scale height by factors of up to two do not
change our results within the statistical noise.

\begin{figure}
\resizebox{\columnwidth}{!}{\includegraphics[angle=0]{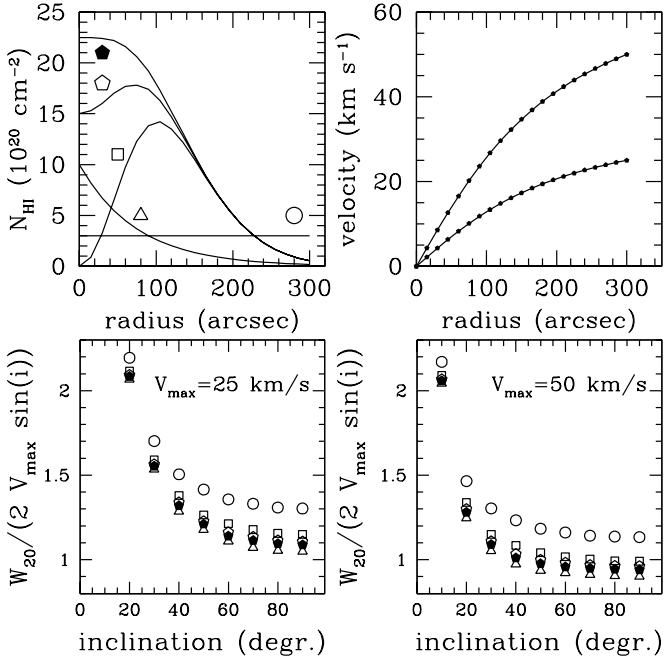}}
\caption{\small Galaxy total HI linewidths as a function of inclination for
two sets of Monte Carlo models. Each set consists of five radial HI
distributions shown at top left, and a rotation curve
shown at top right. Bottom panels show the ratio of
$W_{20}/(2 V_{\rm max} \sin(i)$ as a function of inclination. In the
absence of an isotropic component, this ratio would be constant.
\label{galmod-fig}
} 
\end{figure}

The model is built up of clouds consisting of a large number of
subclouds with no intrinsic velocity dispersion. Cloud positions and
velocities, as well as subcloud velocities are produced by a random
generator in a model data cube according to the specifications of the
gas distribution, rotational velocity, local velocity dispersion,
scale height and projection parameters (position angle and
inclination).  The parameters which define the random realizations are
the number of annuli, the number of clouds per annulus, the number of
subclouds within each cloud and the initialization of the random
generator. As we are interested in the {\it integrated} line profile,
the results presented here are not sensitive to the number of clouds
in a single ring.  The statistical noise in the linewidth $W_{20}$ as
determined from different random realizations of the same model, is of
the order of $0.1\ \rm km\ s^{-1}$.  The models were calculated on a
grid of $5'' \times 5'' \times 4.12\ \rm km\ s^{-1}$, identical to the
sampling of our WSRT observations (Stil \& Israel, in
preparation). The flux of the model in each of the individual velocity
channels yields the model line profiles.  The models contain 20
concentric annuli each, with 5 clouds per $10^{20}\ \rm cm^{-2}$ HI
column density and 1000 subclouds per cloud to define the internal
velocity structure. Experiments with larger numbers of clouds produced
similar linewidths to within $0.1\ \rm km\ s^{-1}$.

\begin{figure*}
\resizebox{\columnwidth}{!}{\includegraphics[angle=0]{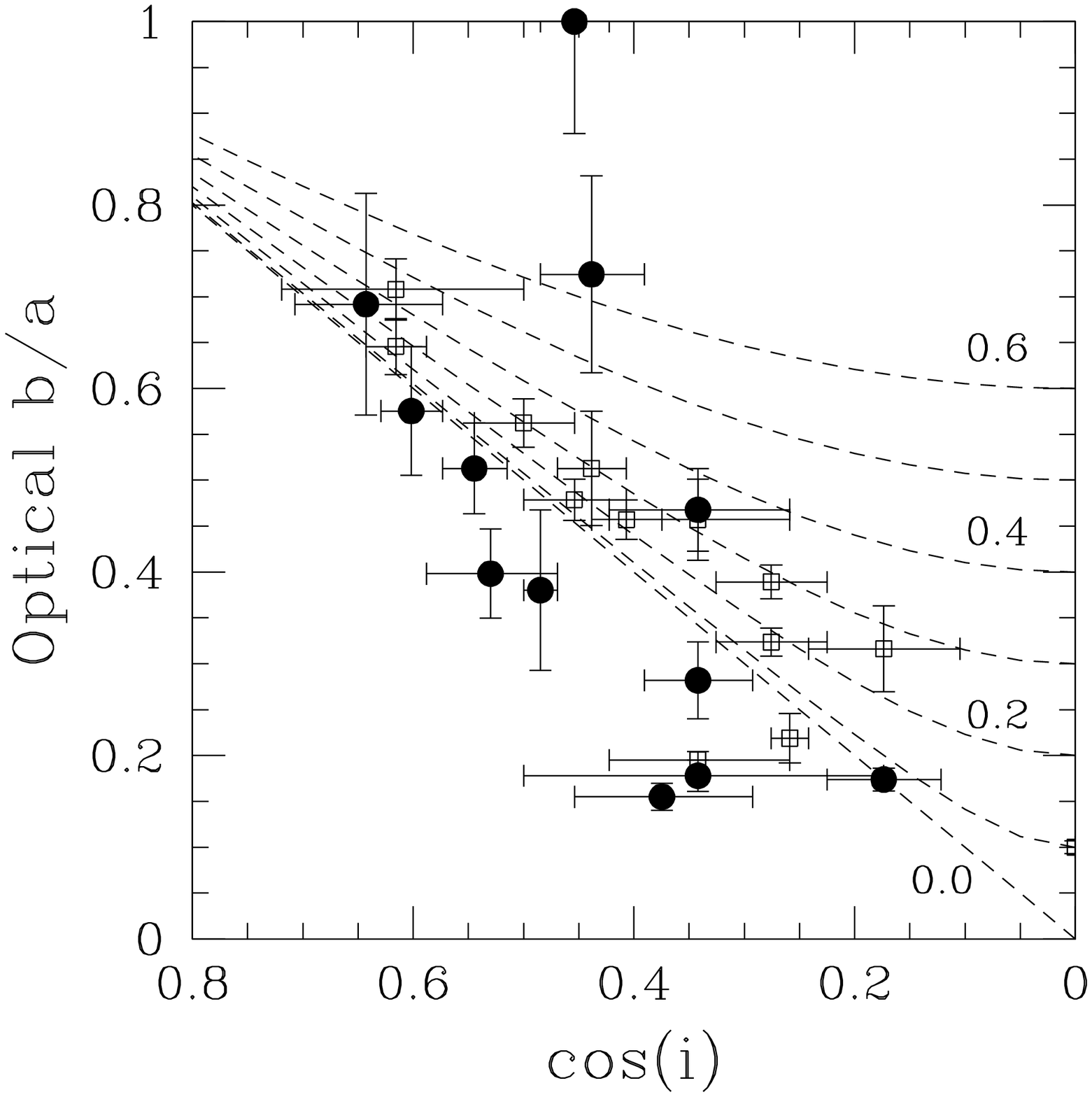}}
\resizebox{\columnwidth}{!}{\includegraphics[angle=0]{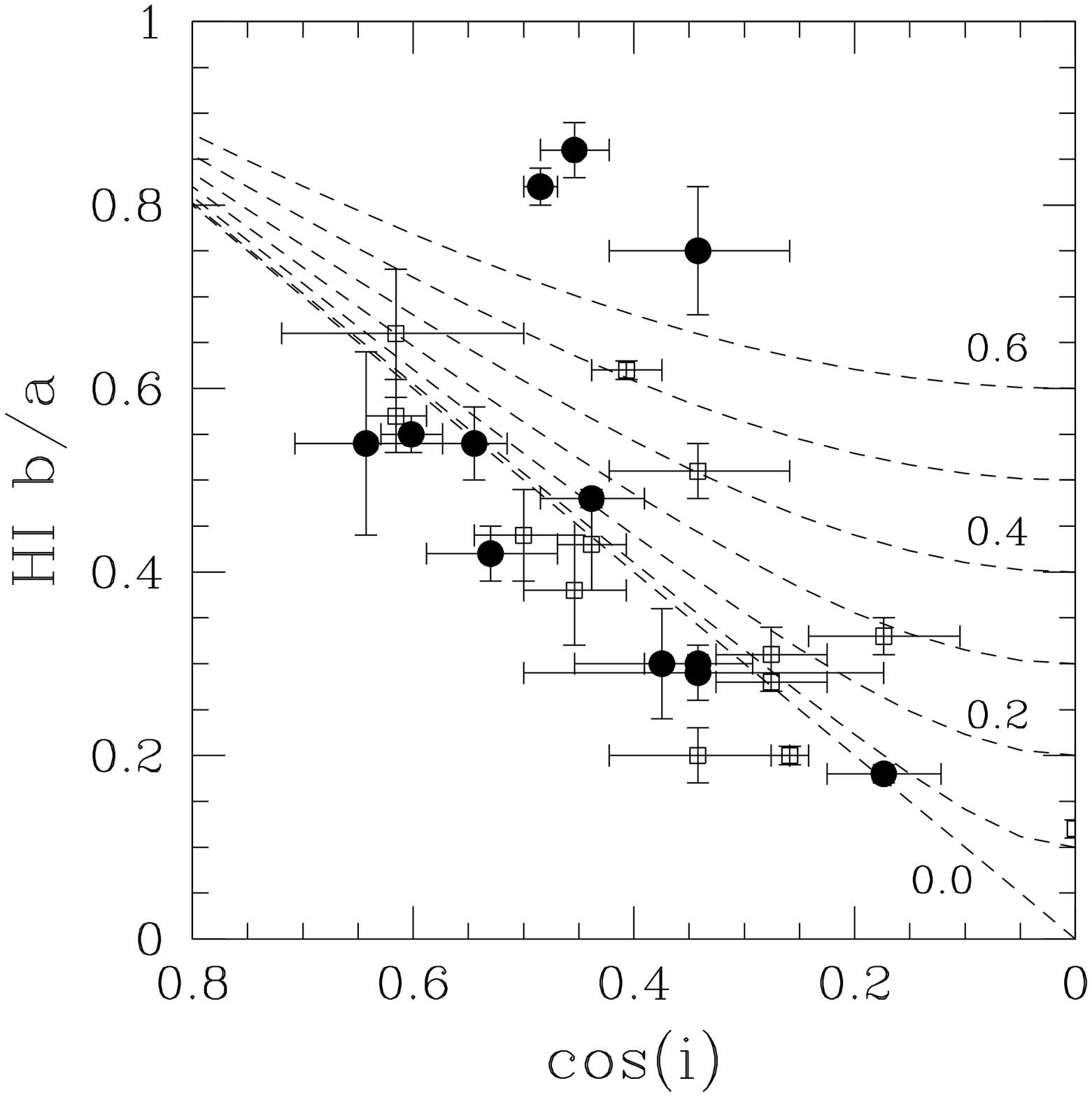}}
\caption{\small Optical axial ratios (top) and HI axial ratios
(bottom) as a function of $\cos(i)$ for the rotation curve sample. Inclination
increases from left to right. Filled circles indicate the dwarfs 
($M_{\rm B}>-16$), open squares the more luminous galaxies. Dashed lines show 
observed axial ratios as a function of $\cos(i)$ according to equation (1) 
for intrinsic axial ratios 0, 0.1, 0.2, $\ldots$, 0.6. 
\label{boveracosi}
} 
\end{figure*}

The rotation curve of an isothermal sphere was used to define the rotation 
velocity as a function of radius. This is an accurate representation
for the shape of the rotation curves of dwarf galaxies (Kravtsov et al. 1998, 
Stil \& Israel 1998). Thus:
$$
v(r)=V_0 {r_c \over r}(1-{\rm arctan}({r \over r_c}))
$$
The core radius $r_c$ and the maximum rotation velocity $V_0$ were
varied independently. A solid body rotation curve corresponds to 
$r_c >> R_{\rm HI}$, a flat rotation curve with $r_c << R_{\rm HI}$. The 
models discussed here have rotation curves with $r_c = 0.5 R_{\rm HI}$ and 
maximum rotational velocities of $25~\rm km\ s^{-1}$ and $50~\rm km\ s^{-1}$ 
respectively. 

All models were calculated with a line of sight velocity dispersion of 
$9~\rm km\ s^{-1}$. This is the mean velocity dispersion in our WSRT sample
of dwarf galaxies.

Five shapes for the radial HI distribution were considered. These are
shown in Fig.~\ref{galmod-fig}.  The profiles represent different
HI column density profiles in our sample.  The shape of the line
profile, and therefore its width at a percentage of the peak, are not
affected by the normalization of the column density.  Therefore, we
treat the rotation curve and the gas distribution independently. 

Each model was evaluated for inclinations 10 to 90 degrees at 10
degree intervals.  Model $W_{20}$ linewidths relative to the projected
maximum rotation velocity are plotted in Fig.~\ref{galmod-fig}. If
the isotropic velocity component can be neglected, the ratio
$W_{20}/(2 V_{\rm max} \sin(i))$ depends only on the gas distribution
and the shape of the rotation curve. Centrally concentrated
distributions, such as an exponential profile (triangles in
Fig.~\ref{galmod-fig}), result in linewidths about 30\% lower than
those reflecting a homogeneous distribution (circles).
  
If the isotropic component (i.e. the line of sight velocity dispersion 
which we kept at $9~\rm km\ s^{-1}$) contributes significantly to the observed
linewidth, the ratio $W_{20}/(2 V_{\rm max} \sin(i))$ increases. This
is illustrated in Fig.~\ref{galmod-fig} which shows the linewidths for 
two different rotation curves. The ratio $W_{20}/(2 V_{\rm max} \sin(i))$ 
is significantly larger for the $V_{\rm max}=25~\ \rm km\ s^{-1}$ rotation 
curve; the difference increases with decreasing inclination.  
The scatter introduced by the HI distribution is approximately 15\%. 
Within this scatter, the approximation $\Delta v_{21}\approx W_{20}/\sin(i)$ 
does not introduce a significant error if the inclination $i>50^{\circ}$ and 
$V_{\rm max} > 25~\rm km\ s^{-1}$. At inclinations between 30 and 50 degrees, 
the increase in $W_{20}/(2 V_{\rm max} \sin(i))$ remains small for the 
models with $V_{\rm max}\approx 50~\rm km\ s^{-1}$, but for those with 
$V_{\rm max}\approx 25~\rm km\ s^{-1}$, only upper limits to 
$\Delta v_{21}$ are obtained.

The models show that a straightforward inclination correction 
$1/\sin(i)$ may be applied to dwarfs with rotation velocities 
as low as $25~\rm km\ s^{-1}$ provided $i>\rm 50^\circ$. 
A more advanced method requires the inclusion of the gas distribution
in the inclination correction, which is beyond the scope of this paper.

The Monte Carlo models, of course, do not include the uncertainty in
the inclination itself, which is difficult to measure. The least
biassed method is to fit the inclination as a free parameter in a
tilted ring analysis of the velocity field. This is not possible if
the rotation velocity is small or if the rotation curve increases
linearly with radius. In those cases, we use
$$
\cos^2(i)={q^2-q_0^2 \over 1-q_0^2} \eqno (2)
$$ 
with the intrinsic axial ratio $q_0=0.15$. Such a small axial ratio
may not be realistic for dwarf galaxies (cf. Staveley-Smith et
al. 1992), but it is a reasonable lower limit to the true axial ratio
of any dwarf galaxy. Inclination corrections $1/\sin(i)$ thus derived
are therefore upper limits. For prolate galaxies, this method would
yield erroneous inclinations. However, most of the observed galaxies
exhibit good alignments of the optical, HI and kinematic major axes,
which suggests that this situation does not apply.

In order to verify the quality of the inclination corrections so obtained,
we have verified that they agree with the kinematic inclinations derived 
for the rotation-curve sample. 
As HI axial ratios are seldom given in the
literature, we measured the axial ratio of the contours near the level
$N_{\rm HI}=2 \cdot 10^{20}\ \rm cm^{-2}$ in HI column density maps
in the references of the rotation curves. The optical axial ratios 
were taken from the Third Reference Catalogue (de Vaucouleurs et al. 
1991). Fig.~\ref{boveracosi} shows the relation between the observed axial
ratio and the kinematic inclination for the rotation curve sample.
The result shows both axial ratios and $\cos(i)$ to be well-correlated, as
expected for flattened systems. The correlation suggests intrinsic
axial ratios $b/a \leq 0.3$. A few dwarfs have observed axial ratios 
slightly smaller than those expected for an infinitely thin system with 
the same inclination. 

Some other dwarfs show departures from the correlation which are consistent
with a larger intrinsic axial ratio. However, deviations from axial 
symmetry and variation of the intrinsic axial ratios $q_0$ have
similar consequences in Fig.~\ref{boveracosi}. The objects with the
most extreme HI axial ratio are DDO~87, DDO~168, and IC~3522. DDO~168 and 
IC~3522 also have optical axial ratios which are approximately half their
HI axial ratio. We nevertheless conclude that the optical and HI axial ratios
are consistent with the kinematic inclination for the majority of 
the galaxies in the RC sample.  

The derived value of $\Delta v_{21}$ is an upper limit for the
slow rotators, as the rotation velocity is small and the inclination
is derived from the observed axial ratio. For the fast rotators, 
$\Delta v_{21}=W_{20}/\sin(i)$, within the limits discussed above and
the inclination is derived from the tilted ring fits.

\section{Results}

\subsection{The Tully-Fisher relation for the two samples}
\label{TF-sec}

\begin{figure}
\resizebox{\columnwidth}{!}{\includegraphics[angle=0]{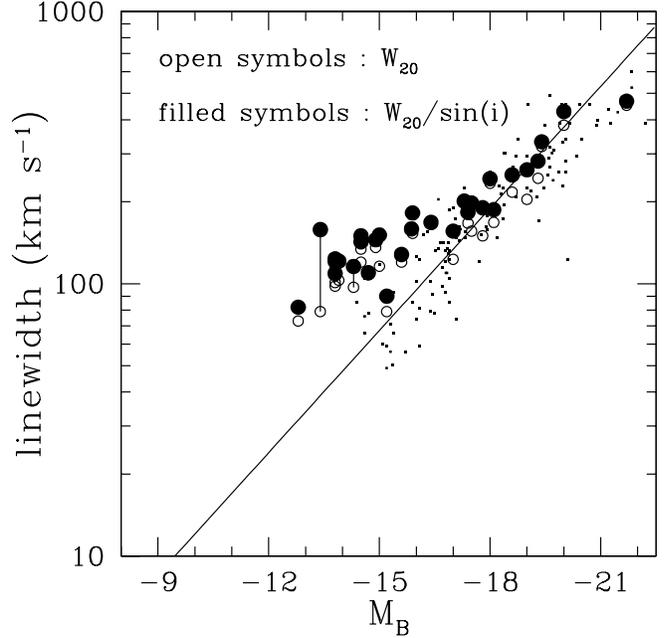}}
\caption{\small Luminosity-linewidth relation for the galaxies in
the rotation curve sample (large symbols). Open circles represent the
measured linewidths before inclination correction; filled circles
show the inclination-corrected linewidths. If the open and filled
circles are connected by a line, the filled circle is an upper limit 
only, as explained in the text. The diagonal line is a fit
of the TF relation by Kraan-Korteweg et al. (1988) to a sample of
nearby galaxies. Small dots are Virgo cluster galaxies used by these 
authors to study the Tully-Fisher relation for different galaxy types.
\label{TF-RCsample}
}
\end{figure}

\begin{figure}
\resizebox{\columnwidth}{!}{\includegraphics[angle=0]{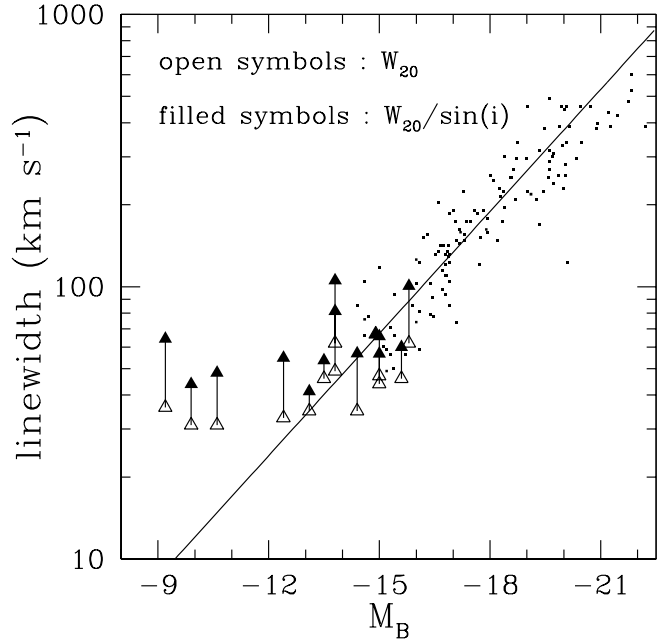}}
\caption{\small As Fig.~\ref{TF-RCsample}, but now for the sample of slow
rotators. The open and filled triangles are connected to emphasize the
range of plausible inclination corrections as explained in the text.
\label{TF-slowrot}
} 
\end{figure}

The relation between $W_{20}$ and $M_{\rm B}$ for the fast rotators
is shown in Fig.~\ref{TF-RCsample}. The reference frame is provided
by the Tully-Fisher relation for a sample of local calibrating galaxies with
luminosity $-16.4 > M_{\rm B} > -21.7$ and a sample of Virgo cluster
galaxies from Kraan-Korteweg et al. (1988).  The open symbols
represent the linewidths before the inclination correction. The
inclination-corrected linewidths are shown as filled circles.

If random motions contribute significantly to the width of the 21 cm
line, the widely derived line of sight velocity dispersion 
$\sigma \approx 10\ \rm km\ s^{-1}$ implies that the linewidth should
be $\Delta v_{21}=W_{20}=36\ \rm km\ s^{-1}$ in the limit of zero 
rotation velocity and isotropic random motions with a gaussian velocity
distribution.

Essentially all galaxies in the rotation-curve sample with relatively high 
luminosities $M_{\rm B} < -17$ follow the Tully-Fisher relation. The most 
luminous sample galaxy (NGC~801) falls within the scatter defined by
the Virgo galaxies. In contrast, the fainter rotation-curve sample 
galaxies deviate from the extra\-polated T-F relation. They either are 
underluminous or have a larger linewidth than expected from this relation.
This is even the case if we use $W_{20}$ linewidths corrected for 
instrumental resolution but not for inclination. We may therefore rule 
out that this deviation is an artifact caused by the inclination
correction. In addition, the maximum rotation velocity of the fast rotators 
is well above $50\rm\ km\ s^{-1}$ and we showed in Section~\ref{model-sec} 
that errors introduced by ignoring the contribution of random motions to the
linewidth are small for such galaxies.

DDO~168 is the only dwarf galaxy in the rotation-curve sample that is 
consistent with the extrapolation of the T-F relation, thereby breaking the
pattern established by the other dwarfs. However, its central HI column 
density $N_{\rm HI}=6.3 \times 10^{21}~\rm cm^{-2}$ is exceptionally high. 
In Section~\ref{model-sec} we have shown that galaxies with a centrally 
concentrated HI distribution have linewidths 20\% to 30\% lower than those 
with a homogeneous HI distribution. If we consider the maximum rotation 
velocity instead of the linewidth, DDO~168 is no longer different from 
the other fast rotators (see Fig.~\ref{TFdisc}). We thus conclude that 
DDO~168 would have a linewidth comparable to those of the other dwarfs, 
if it had a smoother HI distribution.

The luminosity-linewidth relation for the slow rotators is shown in
Fig.~\ref{TF-slowrot}. The open and filled triangles are connected in
order to emphasize that these are lower and upper limits to the true
linewidth $\Delta v_{21}$. The mean linewidth $W_{20}$ of the five
faintest slow rotators with $M_{\rm B}>-13$ is $W_{20}=33.2 \pm
2.3\rm\ km\ s^{-1}$, which is the expected value for a non-rotating
galaxy with a line of sight velocity dispersion of $9\ \rm km\
s^{-1}$. The inclination correction $W_{20}/\sin(i)$ is meaningless
for such objects, as is the extrapolation of the T-F relation itself.
The more luminous slow rotators follow the extrapolation of the
Tully-Fisher relation in the luminosity range $-13 > M_{\rm B} > -16$,
although DDO~75 and DDO~187 (both at $M_{\rm B}=-13.8$) are relatively
far from the T-F relation.  Both are observed at a low inclination and
DDO~75 may in fact be a fast rotator or an intermediate
object. However, the upper limit to the rotation velocity of DDO~187
is $5\rm\ km\ s^{-1}$, which classifies it as a slow rotator if the
inclination from its observed axial ratio is correct

We do not argue that dwarf irregular galaxies fall into two distinct
categories. In fact, we suggest that DDO~168 is an intermediate object.
We checked that the remaining dwarfs in our WSRT
sample (Stil \& Israel, in preparation) for which a complete rotation curve analysis
was not possible, are consistent with the results in Fig.~\ref{TF-RCsample} and 
Fig.~\ref{TF-slowrot}. The uncorrected linewidths $W_{20}$ of these galaxies 
cover the range indicated by the fast rotators and the slow rotators.
The failure of a complete tilted ring analysis was due to the small size,
the small inclination or the near solid-body shape of the rotation curve. 
The four dwarfs added to Table~\ref{sample1} fall into this category.
The RC and SR samples are most likely extremes of a
continuous distribution than two separate groups.

The two samples show that low-luminosity dwarf galaxies deviate systematically
from the extrapolation of the Tully-Fisher relation for luminous
spiral galaxies by being fainter than expected from their linewidth. 
There are no galaxies in the sample for which the converse is true, i.e.
which appear too luminous for their linewidth by a similar amount.
However, Kraan-Korteweg et al. (1988) found that the Im type galaxies in 
their Virgo sample are slightly more luminous than expected from their 
linewidths.

\subsection{Gas content}

\begin{figure}
\begin{minipage}[t]{\columnwidth}
\mbox{}\\
\resizebox{\columnwidth}{!}{\includegraphics[angle=0]{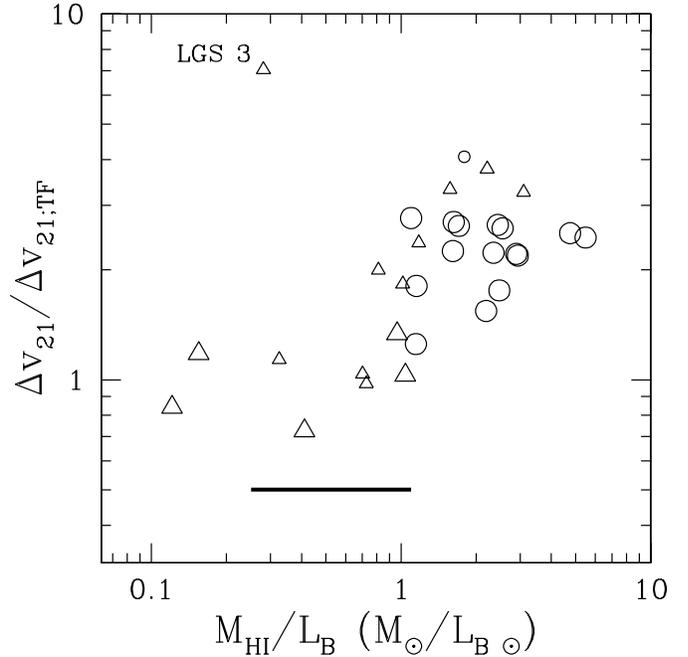}}
\end{minipage}\hfill
\parbox[t]{\columnwidth}{
\caption{ Ratio of the inclination-corrected linewidth $\Delta v_{21}$ 
and the linewidth inferred from luminosity and the T-F relation, as a
function of gas content $M_{\rm HI}/L_{\rm B}$.  Circles indicate the fast
rotators, triangles the slow rotators.  Objects with inclinations
less than $50^\circ$ are represented by small symbols. The horizontal bar 
marks the range of $M_{\rm HI}/L_{\rm B}$ for Sd-Im type galaxies
in the sample of Rhee (1996).
\label{MHoverLB-fig}
}} 
\end{figure}

In Fig.~\ref{MHoverLB-fig} we show the ratio of the actual inclination
corrected linewidth to the linewidth predicted by the T-F relation,
as a function of gas content measured by the ratio $M_{\rm HI}/L_{\rm B}$.
Gas-rich dwarf galaxies with $M_{\rm HI}/L_{\rm B} > 1~\rm 
M_{\odot}/L_{B\odot}$ deviate increasingly from the T-F relation with 
increasing gas content. The fast rotators are all gas-rich. 
The gas-rich ``slow rotators'', all objects with a small inclination ($i<50^{\circ}$), 
exhibit the same behavior. Note that $\Delta v_{21}$ values for these 
galaxies are upper limits to the true linewidth for these galaxies, so that
the apparent offset between triangles and circles is not meaningful.
Slow rotators less rich in gas ($M_{\rm HI}/L_{\rm B} <$ 
$1~\rm M_{\odot}/L_{B \odot}$) are consistent with the T-F relation.

This correlation with a parameter which is independent of the
distance, shows that the large deviations from the T-F relation found
in Section~\ref{TF-sec} cannot be caused by systematic errors in the
assumed distance. We have included the three faintest slow rotators in
Fig.~\ref{MHoverLB-fig} although the extrapolation of the T-F
relation is no longer valid at such low luminosities as it predicts
linewidths smaller than expected for HI velocity dispersions of $10\rm
\ km\ s^{-1}$. This is in fact the cause of the anomalous location of
the very faint dwarf galaxy LGS~3 in the diagram.

Finally, we have marked in Fig.~\ref{MHoverLB-fig} by a horizontal bar
the $M_{\rm HI}/L_{\rm B}$ range of Sd-Im galaxies in the sample of Rhee 
(1996). The fast rotators are mostly outside this range. This explains why 
large deviations from the T-F relation did not appear in previous studies: 
fast rotators were not represented in those samples.

\subsection{Virial and baryonic mass}
\label{MvirMbar-sec}

\begin{figure*}
\resizebox{\textwidth}{!}{\includegraphics[angle=0]{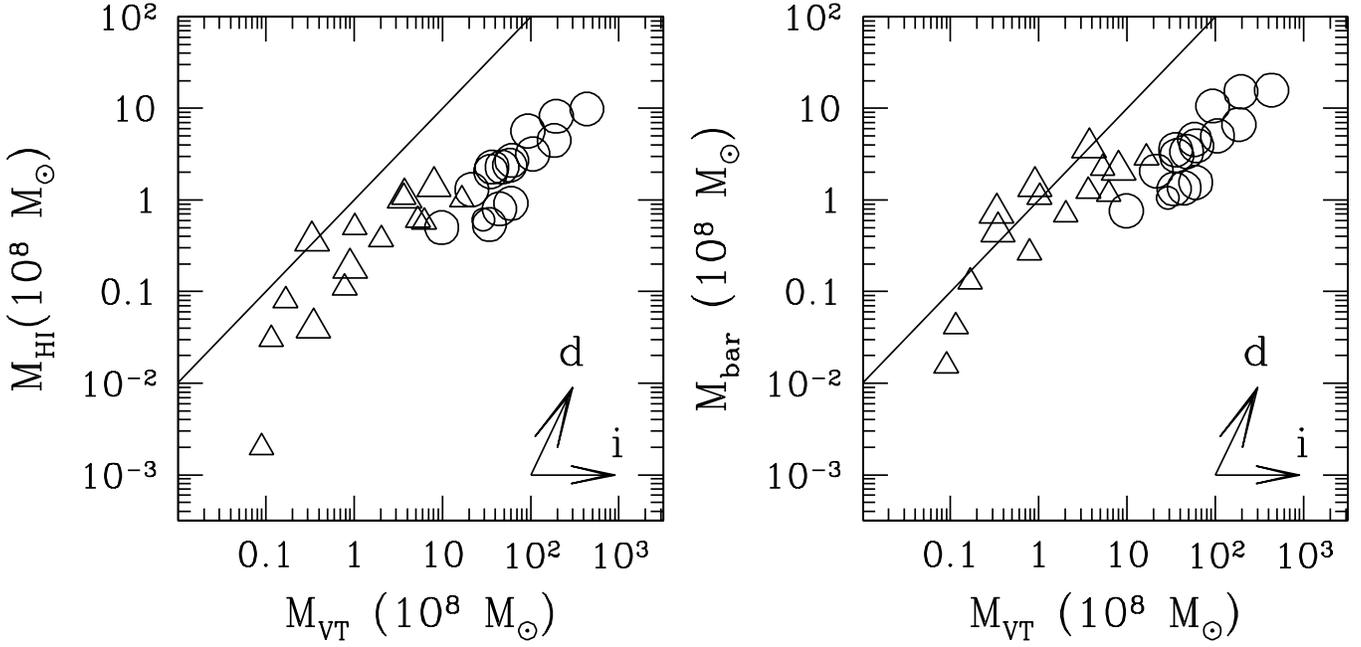}}
\caption{\small {\it Left:\/} Relation between virial mass and
HI mass for the dwarf galaxies in Fig.~\ref{MHILB-fig}. 
The solid line is the line of equal mass. Arrows indicate direction and 
magnitude of errors corresponding to uncertainties of factors of three
distance (labeled ``d'') and in $\sin(i)$ (labeled ``i''), assuming the 
galaxy is transparent.
{\it Right:\/} Relation between virial mass and baryonic
mass. The slow rotators are dominated by baryonic matter, whereas the
fast rotators are dominated by dark matter (Section~\ref{MvirMbar-sec}).
\label{MvtMlum-fig}
} 
\end{figure*}

The total mass within the HI radius $R_{\rm HI}$ follows from the virial
theorem (Lo et al. 1993):
$$
M_{\rm VT}=\beta R_{\rm HI} G^{-1} (V^2_{\rm rot}+ 3 \sigma^2)   \eqno (3)
$$
where $\sigma$ is the observed line of sight velocity dispersion of the HI.
Although the term $V^2_{\rm rot}$ depends strongly on the inclination, the 
importance of the isotropic random motions in the slow rotators reduces the
errors introduced by the uncertain inclination. 
The parameter $\beta$, of order unity, depends on the actual but poorly 
known distributions of HI and dark matter. 
Following Lo et al. (1993), a constant value $\beta=5/3$ is adopted,
corresponding to a homogeneous HI distribution in a halo with a
constant density. Although this value does not apply to all of
the galaxies discussed here, the systematic error introduced by this 
is less than a factor 2. This is small compared to the mass range which
covers four orders of magnitude.

We adopt the radius of the last measured point of the rotation curve as
the radius $R_{\rm HI}$. The uncertainty in the parameter $\beta$ is considerably
larger than the uncertainty introduced by this approximation. 

The baryonic mass is the sum of the gas mass and the stellar mass. 
The contribution of molecular gas to the overall gas mass in dwarf 
irregular galaxies is believed to be small. Israel (1997) found 
the ratio of molecular to atomic hydrogen mass to be about 0.2 for 
a sample of seven nearby dwarfs. Consequently, we ignore the 
possible contribution of $\rm H_2$. 
The baryonic mass is then 
$$
M_{\rm bar}= (1+{M_{\rm He}\over M_{\rm H}})M_{\rm HI} + \Bigl({M \over 
L_{\rm B}}\Bigr)_* L_{\rm B} \eqno (4) 
$$
where we adopt a ratio of helium to hydrogen 
$M_{\rm He}/M_{\rm H}=0.3$. 

The mass-to-light ratio of the stellar population is calculated from  
$(B-V)$ colours and the relation (Bottema 1997)
$$
\Bigl({M \over L_{\rm B}}\Bigr)_*=1.93 \cdot 10^{0.4(B-V)} -1.88 \eqno (5)
$$
The normalization of this relation depends on the poorly known
low-mass end of the stellar IMF. However, the strong dependence of the
mass-to-light ratio on the colour of the stellar population is taken
into account. For a typical object in our sample with $(B-V)=+0.4$,
the stellar mass-to-light ratio is 0.91. For the bluest dwarf
galaxies, $(B-V)\approx 0$, the derived stellar mass-to-light ratio
is very small and much more uncertain. Broeils (1992) found a maximum
stellar mass-to-light ratio restricted by the rotation curve in the
range 0.5 - 3 for 6 galaxies of type later than Sd.

Fig.~\ref{MvtMlum-fig} shows the relations between HI mass and
virial mass, and between baryonic mass and 
virial mass in the two samples. The difference in virial mass between
the slow rotators and the fast rotators is implicit in the sample
selection.  Although both samples show a correlation between HI mass and 
virial mass, the slopes are poorly constrained due to the relatively 
large scatter. A least-squares fit to both samples, excluding the
faintest 3 dwarfs with $M_{\rm B}>-12$, results in the relations
$$
{\rm log}\ M_{\rm HI} = -0.41 + 0.55\ {\rm log}\ M_{\rm VT} \pm 0.27
$$
The inverse regression yields
$$
{\rm log}\ M_{\rm VT} = 0.93 + 1.38\ {\rm log}\ M_{\rm HI} \pm 0.42
$$
For the relation between baryonic mass and virial mass, we have likewise
$$
{\rm log}\ M_{\rm bar} = -0.14 + 0.41\ {\rm log}\ M_{\rm VT} \pm 0.26  
$$
$$
{\rm log}\ M_{\rm VT} = 0.66 + 1.59\ {\rm log}\ M_{\rm bar} \pm 0.51 
$$
All masses are in units of $10^8\ M_\odot$. The errors are the r.m.s.
residuals of the fits. The slope of the fit is sensitive to the
treatment of the parameter $\beta$. If $\beta$ is systematically
smaller for the fast rotators, the slope of the relation between 
${\rm log}\ M_{\rm HI}$ and ${\rm log}\ M_{\rm VT}$ is steeper as a function 
of $M_{\rm VT}$ is steeper than these fits indicate. However, the slope is 
significantly less than unity for all plausible values of $\beta$. 
Therefore, the lower-mass systems have a larger fraction of their virial 
mass in gaseous form. For the fast rotators, the virial mass is approximately
an order of magnitude larger than the baryonic mass.
The fast rotators have velocity gradients within 
one kpc from the centre of typically $30\ \rm km\ s^{-1}\ kpc^{-1}$, about 
a factor of three larger than those of the slow rotators. This implies 
central mass densities for the fast rotators approximately an order of
magnitude higher than for the slow rotators.

Within the uncertainties introduced by the stellar mass-to-light ratio
and the assumptions made to calculate the virial mass, the baryonic
mass equals the virial mass for the slow rotators in the luminosity
range $-13>M_{\rm B}>-16$. 

\begin{figure}
\resizebox{\columnwidth}{!}{\includegraphics[angle=0]{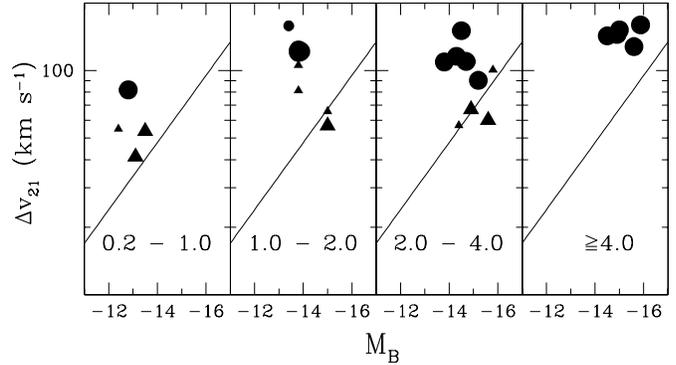}}
\caption{ Position of dwarf galaxies binned by baryonic mass in
the luminosity-linewidth diagram. Size and shape of the symbols
are as in Fig.~\ref{MHoverLB-fig}. Diagonal line represents the
Tully-Fisher relation. The ranges in baryonic mass are given in each
panel in units of $\rm 10^8\ M_\odot$.
\label{MbarTF-fig}
} 
\end{figure}

Fig.~\ref{MbarTF-fig} shows the position of the fast rotators and
the slow rotators relative to the T-F relation as a function
of baryonic mass. Galaxies from both samples with similar baryonic 
mass form well-ordered groups in Fig.~\ref{MbarTF-fig}.
At similar baryonic mass, the slow rotators tend to be
more luminous than the fast rotators because their stellar mass is a larger
fraction of the baryonic mass (see also Fig.~\ref{MHoverLB-fig}).
It is clear from Fig.~\ref{MbarTF-fig} that the baryonic mass
increases along the T-F relation. Qualitatively, this result 
does not depend on the assumed stellar mass-to-light 
ratio, but the precise direction in which baryonic mass 
increases is sensitive to its adopted value.

There is no independent evidence for an
evolutionary connection between fast rotators and slow rotators with the
same baryonic mass.  However, if mass loss due to galactic winds or
mergers can be ignored, baryonic mass is conserved in the process of
star formation.  On the other hand, if a significant amount of gas is
expelled in a starburst phase, the total baryonic mass decreases.
Therefore, this result places constraints on the evolution of galaxies
through star formation in the luminosity-linewidth diagram. We will
return to this point.

\section{Discussion}

\subsection{Reality of the deviation}

In Section~\ref{TF-sec} we showed that faint dwarf galaxies in
general do not follow the extrapolation of the T-F relation established
for luminous galaxies. This effect decreases with luminosity and
disappears into the scatter in the blue-TF relation near $M_{\rm B}=-16$.  
However, a well-defined, surprisingly homogeneous, subclass of faint dwarf
galaxies (sample 2) appears to obey the Tully-Fisher relation. Most other
dwarf galaxies are less luminous than expected from their
linewidth. The present samples do not contain dwarf galaxies which are 
overluminous to similar degrees. The largest observed deviation is 
three magnitudes (factor 15) in luminosity or a factor of two in
linewidth. Before we further discuss the result, we consider the 
possibility that this effect is the result of systematic errors.

There is no systematic error in the linewidth $\Delta v_{21}$ which
could have produced this result. The difference is large, even if we use
directly measured linewidths, corrected for instrumental resolution
but not for inclination. 
If the velocity of the flat part of the rotation curve is considered as
the velocity parameter of the T-F relation,
the linewidth of galaxies with a rotation curve which rises at the last 
measured point is expected to be {\it smaller} than its T-F value. 
A similar argument can be made for the HI distribution, as $\Delta v_{21}$
cannot exceed the velocity of the HI at the edge of the galaxy, but it can be 
considerably less if the galaxy has a central concentration of HI. 

This leaves the possibility of systematic effects along the luminosity axis.  
If our result is to be ascribed to distance errors, we must have been
underestimating distances systematically by up to a factor of four. 
Moreover, this systematic distance error must then be luminosity-dependent, 
because the magnitude of the deviation decreases as the linewidth increases. 
The argument in Section~4.2, involving high but distance-independent
$M_{\rm HI}/L_{\rm B}$ ratios (cf. Fig.~\ref{MHoverLB-fig}) also
argues against this possibility.

We conclude that the extrapolation of the Tully-Fisher relation for
luminous galaxies to lower luminosities is not representative for the
majority of dwarf galaxies. Dwarf galaxies are on average less
luminous than predicted from the width of the 21 cm line and the T-F
relation. Such a large deviation from the T-F relation was, in fact, 
first noticed by Carignan \& Beaulieu (1989) for the dwarf galaxy 
DDO~154 and later by Meurer et al. (1996) for NGC~2915. Both objects 
have been included in the present sample.

Matthews et al. (1997) found a smaller 
deviation of 1.3 magnitudes from the Tully-Fisher relation for a 
large sample of extreme late-type galaxies in the luminosity range 
$-14.5>M_{\rm B}>-16$. This is consistent with the deviation in our 
RC sample in the same luminosity range.  Matthews et al. (1997) also 
concluded that the baryonic correction was insufficient to
place their objects on the T-F relation.

\subsection{Origin of the deviation}
\label{TFdisc-sec}

The basic shape of the T-F relation can be derived from simple
arguments (see also Persic 1993 and Zwaan et al. 1995).
Assume that the optical radial surface brightness profile is
exponential with a central surface brightness $\Sigma_0$ and a radial
scale length $h$. The luminosity is $L \sim \Sigma_0 h^2$.
This does not exclude a considerable vertical scale height. 

\noindent
The mass within radius $R$ follows from the virial theorem : 
$M \sim R (v^2+\sigma^2)$.  If the width of the 21 cm HI line is used, 
$R$ is a fixed point in the rotation curve, unless the rotation curve is 
flat. Usually, a constant value for the quantity $R/h$ is 
assumed, and $h$ is
eliminated to derive the T-F relation. However, both $h$ and
$R$ are fixed by the data and $h \sim R$ is a dubious
assumption. If the ratio $R/h$ is explicitly included, the relation 
between luminosity and velocity is
$$
L \sim {\Bigl({R \over h}\Bigr)^2 \over \Sigma_0 \Bigl({M \over L}\Bigr)^2}\ (v^2+\sigma^2)^2 \eqno (6)
$$
This expression deviates less than 10\% from the familiar 
$L \sim v^4$ for $v > 4 \sigma$.  Note that 
$R/h$ and $M/L$ are correlated because $M$ is the mass within 
radius $R$.

However, in classic (non-LSB) spiral galaxies with a flat rotation
curve, the mass-to-light ratio of the stars can be scaled to explain
the inner, rising part of the rotation curve with the stellar
mass. The rotation curve of a thin exponential disc is maximal at 2.2
scalelengths (Freeman 1970). Therefore, the linewidth measures the
velocity at $R=2.2\ h$, where the luminosity is related to the mass
through the mass-to-light ratio within the stellar disc.

\begin{figure}
\resizebox{\columnwidth}{!}{\includegraphics[angle=0]{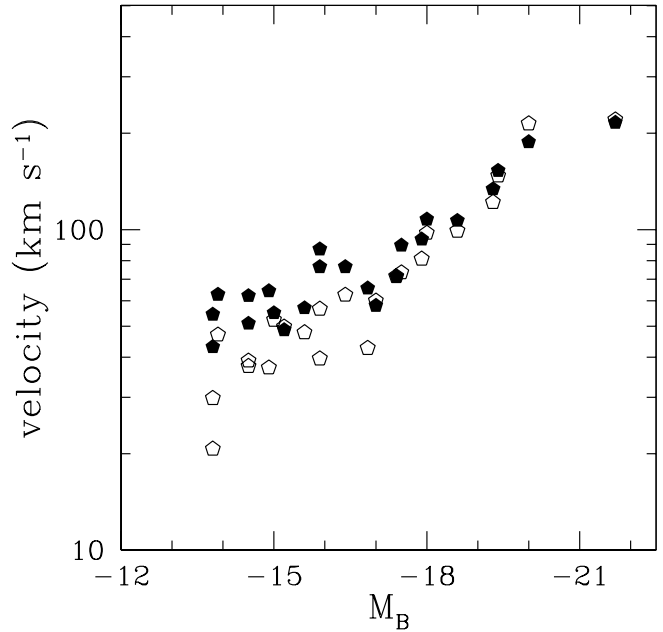}}
\caption{ Tully-Fisher relation for the RC sample in the form
of the maximum rotation velocity (filled symbols) and the rotation
velocity at 2.2 exponential scalelengths (open symbols) as a function
of $M_{\rm B}$.
\label{TFdisc}
} 
\end{figure}

Zwaan et al. (1995) showed that LSB galaxies follow the same T-F
relation as high-surface-brightness (HSB) galaxies.  They
inferred the relation $\Sigma_0\sim (M/L)^{-2}$ as a result of
the identical zeropoint of the T-F relations for both high and low
surface-brightness galaxies. 
The inferred correlation between surface-brightness and mass-to-light
ratio is confirmed by a detailed comparison between an LSB galaxy and
an HSB galaxy at the same position on the T-F relation (De
Blok \& McGaugh 1996). The LSB galaxies considered by Zwaan et
al. (1995) have luminosities $M_{\rm B}<-15.5$. The ratio between
the velocity at 2.2 exponential scalelengths and the velocity at the
last measured point of the rotation curve is $0.91 \pm 0.11$ (data
from De Blok 1997).  Therefore, the 21 cm HI linewidth is a good
representation of the rotation velocity at 2.2 exponential
scalelengths. The assumption $R=2.2\ h$ is valid for the LSB
galaxies in the sample of Zwaan et al. (1995), even though the stellar
disc cannot account for the rotation velocity at this point
(De Blok 1997). Therefore, the relationship
$\Sigma_0\sim (M/L)^{-2}$ inferred by Zwaan et al. (1995) is not affected.
This also shows that the relative amplitude of the stellar (partial)
rotation curve to the halo (partial) rotation curve is not a critical
factor in determining whether a galaxy is on the T-F relation. 
Therefore, the
deviation of dwarf galaxies from the T-F relation should
not be interpreted as a breakdown of the so-called ``disc-halo
conspiracy''.

Thus, the observation that the Tully-Fisher relation for luminous
LSB galaxies displays no offset with respect to the T-F
relation for HSB galaxies, implies that the product $\Sigma_0 (M/L)^{2}$ 
is constant. The observed deviation of the fast rotators is then
interpreted as a violation of the assumption $R \sim h$.  

The turnover radius of a dwarf galaxy rotation curve can be much
larger than the scale length of the optical disc. A detailed analysis
of the rotation curves is beyond the scope of this paper, but this can
be checked directly for the objects for which mass models extist in
the literature (see Table~\ref{sample1}). Therefore, the mass measured
with the linewidth includes a variable amount of dark matter,
depending on the distribution of the HI. This suggests that the
deviation of dwarf galaxies from the T-F relation found in
Section~\ref{TF-sec} should disappear if the velocity at $R=2.2\ h$ is
considered. Fig.~\ref{TFdisc} shows the maximum velocity (filled
symbols) and the velocity ar $R=2.2\ h$ (open symbols) as a function
of luminosity for the rotation curve sample. As expected, the maximum
rotation velocity closely follows the behaviour of the linewidth. As
noted before, in this diagram DDO~168 is no longer exceptional. The
deviation of the velocity at 2.2 exponential scale lengths is much
less, but small deviations remain for the faintest objects.
This confirms the importance of the factor $R/h$ in equation (6).

\subsection{No evolution}

Although the high $M_{\rm HI}/L_{\rm B}$ ratio of the fast rotators
implies that they have converted relatively little gas into stars,
there is no evidence that the fast rotators are presently evolving
towards the Tully-Fisher relation. If the star formation rate $\dot{M}_*$ is
sufficiently low, old stars are replaced by new stars at a constant,
or even diminishing, luminosity.  If the stellar rejuvenation
timescale $\tau = M_*/\dot{M}_*$ is of the order of the lifetime of the
presently most luminous stars, the luminosity of the galaxy does not
increase as a result of star formation.

For a dwarf galaxy with luminosity $M_{\rm B}=-14.5$ 
($L_{\rm B}\approx 10^8\ L_{\rm B,\odot}$), and a star formation rate 
$\dot{M}_*=0.01\ M_\odot\ \rm yr^{-1}$, the rejuvenation timescale is
$\tau=10^{10}$ years, i.e. a Hubble time. The star formation rate can 
barely keep up with the fading of the stellar population, illustrating 
that a large reservoir of gas does not mean that a galaxy will be more 
luminous in the future.
 
However, the observation that the most gas-rich dwarfs are presently
underluminous relative to the Tully-Fisher relation justifies a discussion 
of the possibility that these objects will increase their brightness and 
become normal Tully-Fisher galaxies in the future.

\subsection{Luminosity evolution (the baryonic correction)}

The most extreme dwarf galaxies are approximately 2.5 to 3 magnitudes
fainter than the T-F luminosity corresponding to their linewidth.  Do
they contain sufficient amounts of HI to form the stars required to
turn them into Tully-Fisher galaxies?  We consider a gas-rich
dwarf galaxy ($M_{\rm HI}/L_{\rm B}=3$), which is two magnitudes
fainter than expected from the T-F relation. If all HI is converted
into stars with a stellar mass-to-light ratio $(M/L_{\rm B})_*=1$, the
galaxy is still a magnitude too faint. This potential luminosity is sometimes 
called the baryonic correction. Thus, the large luminosity
deficiency can only be repaired by creating a stellar population with
a very low mass-to-light ratio, which is consequently blue and evolves
on a short timescale.

\begin{figure}
\begin{minipage}[t]{9cm}
\mbox{}\\
\resizebox{\columnwidth}{!}{\includegraphics[angle=0]{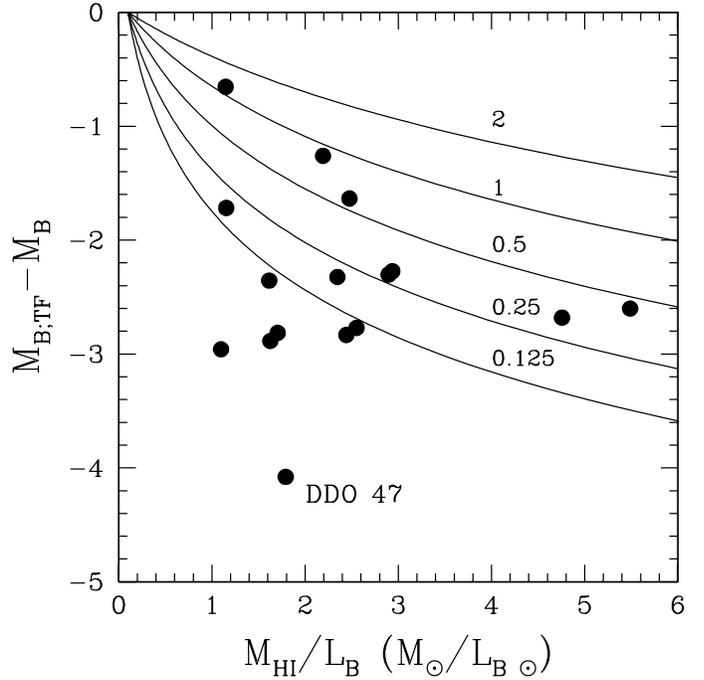}}
\end{minipage}\hfill
\caption{\small Deviation of fast rotating dwarfs from the T-F
relation as a function of HI gas content. The curves indicate the
change in luminosity as a function of initial ratio $M_{\rm HI}/L_{\rm
B}$, if the HI gas is converted into stars to produce a final ratio
$M_{\rm HI}/L_{\rm B} = 0.1$, for five possible values of the stellar
mass-to-light ratio. A calibration of the T-F relation corresponding
to $H_0=75\ \rm km\ s^{-1}\ Mpc^{-1}$ would shift the points upward by
0.6 magnitudes.  The two galaxies with $M_{\rm HI}/L_{\rm B}>4$ are
DDO~154 and NGC~2915.  The inclination of DDO~47, and therefore its
distance from the T-F relation, is very uncertain.
\label{MHILB-fig}
}
\end{figure}

If the mass presently in molecular hydrogen can be neglected (Israel
1997), the HI gas is the reservoir for star formation.  The maximum
luminosity gain to be achieved corresponds to instantaneous conversion
of all HI into stars with a certain mass-to-light ratio. More gradual
star formation would produce less luminosity as previous generations
of stars fade in the meantime.

If a mass $M$ of gas is converted into stars to obtain a small HI
mass-to-blue-light ratio $[{M_{\rm HI}/L_{\rm B}}]_{0}$, we have
$$
{M_{\rm HI}-M \over L_{\rm B} + b M}=[{M_{\rm HI} \over L_{\rm B}}]_{0}
$$ 
with $b^{-1}$ the mass-to- light ratio of the stars. 
Solving for $b M$ yields the relative increase in luminosity 
$$
{b M \over L_{\rm B}} = {b \over 1 + b [{M_{\rm HI} \over 
L_{\rm B}}]_{0}}({M_{\rm HI} \over L_{\rm B}}-[{M_{\rm HI} \over L_{\rm B}}]_{0}) \eqno (7) 
$$
This ratio is shown graphically in Fig.~\ref{MHILB-fig} as a
function of $M_{\rm HI}/L_{\rm B}$ for five values of the mass-to-light 
ratio $1/b$ of the stellar population. The inferred mass-to-light ratio varies 
between unity and less than 1/8 in Fig.~\ref{MHILB-fig},
but it is sensitive to the adopted calibration of the
Tully-Fisher relation. If the zero point of the calibration is shifted
from $H_0=57\ \rm km\ s^{-1}\ Mpc^{-1}$ to $H_0=75\ \rm km\ s^{-1}\ Mpc^{-1}$, 
the distance from the TF relation decreases
by 0.6 magnitudes. In that case, stellar mass-to-light ratios up to
1/4 would allow most of the fast rotators to bridge the gap. According
to equation (5), stellar mass-to-light ratios between 1/8 and 1/4 translate 
into (B-V) colours between 0 and 0.1,
which is considerably bluer than the mean (B-V) of the two samples
discussed here. If the normalization of equation (5) yields
stellar mass-to-light ratios systematically too low, as suggested by
the maximum stellar disc hypothesis, the inferred colours of the
stellar population would be even more blue. Therefore, these low stellar
mass-to-light ratios are unrealistic.

Observational evidence against luminosity evolution to the T-F
relation at constant linewidth is provided by
Fig.~\ref{MbarTF-fig}. The baryonic mass of a galaxy on the T-F
relation exceeds the mass of a fast rotator with the same linewidth by
a factor $\sim3$. This implies that the fast rotators must contain
even more gas, which is not detected in single dish measurements of
the 21 cm line. However, if the mass-to-light ratio of the stellar
population is much smaller than the normalization of Bottema (1997),
the lines of constant baryonic mass would be more horizontal and this
problem does not occur. Although very small mass-to-light ratios can
increase the baryonic correction, and make it consistent with the
distribution of galaxies with similar baryonic mass in the
luminosity-linewidth diagram, the blue colours expected for these low
mass-to-light ratios are not observed.  We therefore conclude that the
evolution to the T-F relation at a constant linewidth is unlikely.

\subsection{Evolution in linewidth and luminosity}

An important constraint for evolution towards the Tully-Fisher
relation is in fact given by the same Fig.~\ref{MbarTF-fig}. If we
assume that baryonic mass is conserved in the process, evolution in
both linewidth and luminosity is consistent with the observed
distribution of galaxies with identical baryonic mass in the
lumino\-sity-linewidth diagram. If a significant amount of gas,
in the order of 50\% or more, is lost due to a galactic wind,
the distribution of baryonic mass in Fig.~\ref{MbarTF-fig} also
implies a decrease in linewidth.
  
As the rotation curve is determined primarily by the dark halo, the
linewidth can be decreased by transporting gas to the
centre, e.g. as the result of tidal interaction with another galaxy.
The luminosity is expected to increase as a result of star formation
in the high-density gas in the centre.

We test this possibility by assuming that the fast rotators had a low
star formation rate $\dot{M}_{*,1}$ for a long period of time, $t_1$,
similar to LSB galaxies (De Blok 1997), followed by shorter period
$\Delta t$ with a high star formation rate $\dot{M}_{*,2}$ as a result of
the concentration of gas towards the centre. Before the starburst, the
luminosity of the galaxy is $L_{\rm B}\approx\dot{M}_{*,1}
t_1/(M/L_{\rm B})_{*,1}$. Similarly, the extra luminosity of the stars
formed in the burst, is $\Delta L_{\rm B} \approx \dot{M}_{*,2} \Delta
t/(M/L_{\rm B})_{*,2}$. We allow for a difference in stellar
mass-to-light ratio between the younger and the older stellar population.
The ratio of these luminosities is
$$
{\Delta L_{\rm B} \over L_{\rm B}} = {\dot{M}_{*,2} \over \dot{M}_{*,1}} {\Delta t \over t_1} \mathcal R \eqno (8)
$$
where $\mathcal R$ is the ratio of the mass-to-light-ratios of the old
and the young stellar population. If the linewidth of a galaxy with a solid
body rotation curve is decreased by a factor f, the mean gas surface density 
increases by a factor $f^{-2}$.
The ratio of the star formation rates follows from the Schmidt
(1959) law with an exponent $N=1.4$ (Kennicutt 1998):
${\dot{M}_{*,2}/\dot{M}_{*,1}}=f^{-2.8}$.  Assume $t_1=10^{10} \rm
yr$, $\Delta t=10^8~\rm yr$ (i.e. of the order of the
of the orbital timescale of the gas), $(M/L_{\rm B})_{*,1}=1$ and $(M/L_{\rm
B})_{*,2}=0.1$.  With these assumptions, equation (8) reduces to
$$
{\Delta L_{\rm B} \over L_{\rm B}} = 0.1 f^{-2.8}
$$
For $f=0.5$ we find $\Delta M_{\rm B}=-0.6$ magnitudes. Variation of
the time of the burst or the mass to light ratio of the stars formed
in the burst by a factor of 3 changes $\Delta M_{\rm B}$ by
approximately 0.5 magnitudes.

The combination of a decrease in linewidth by a factor 0.5 and 
an increase in luminosity of 0.6 magnitudes is consistent with  
the distribution of galaxies with similar baryonic mass in the 
luminosity-linewidth diagram (Fig.~\ref{MbarTF-fig}).

This mechanism provides both the decrease in linewidth and the higher
star formation rate required to move towards the T-F relation.  When
the baryonic mass becomes more centrally concentrated, the dynamical
importance of the halo decreases, and the central galaxy is less
dominated by dark matter. The gas from the outer galaxy is expected to
be metal-poor, thus providing the possibility of a low-metallicity
burst in a galaxy which experienced chemical enrichment in the past.

However, some difficulties remain in this picture. The inner slope of
the rotation curve of the fast rotators is considerably larger than
that of the slow rotators, implying a higher central mass density in
the fast rotators. The published mass models show that this is is
mainly due to the dark halo. The formation of a central concentration
of gas would enhance the density further.  Therefore, a fast rotator
cannot become a slow rotator similar to those of the present SR
sample, unless the density profile of the dark halo is also affected.

We summarize that in principle, there is no reason to suspect that 
the T-F relation which appears universal for large spiral galaxies,
is applicable to dwarf galxies. Therefore, it remains to be shown
that the deviating dwarfs evolve towards the T-F relation.  
The primary reason to suspect evolution towards the T-F relation
is the correlation between $M_{\rm HI}/L_{\rm B}$ and distance
from the T-F relation. However, the baryonic correction,
i.e. evolution at a constant linewidth, is inconsistent with the 
observed large deviations (Matthews et al. 1997 and this work).
This leaves the possibility that the T-F relation is applicable only
to large spirals and {\it some} dwarf galaxies (the slow rotators),
or that evolution in both linewidth and luminosity occurs. 
Circumstantial evidence for the latter is provided by the observed
ordered distribution of baryonic mass in the luminosity-linewidth 
diagram.

\section{Conclusions.}
\label{conclusions}

1. Rotationally supported dwarf galaxies and predominantly pressure
supported dwarf galaxies exist together in the luminosity range $-13 >
M_{\rm B} > -16$.

2. The pressure supported dwarf galaxies (the slow rotators) follow
the extrapolation of the T-F relation for luminous galaxies. The
rotationally supported dwarfs, (the fast rotators) either are systematically
too faint or systematically rotate faster than expected from the T-F 
relation.  There is no sharp dichotomy between these extremes. 

3. The fast rotators have a larger fraction of gas relative to stars
than the slow rotators. Therefore, the fast rotators are less evolved
in terms of star formation than the slow rotators. 
The correlation of gas content with distance from the T-F
relation suggests that evolution in the luminosity-linewidth diagram
may occur. 

4. Within the HI distribution, the dynamics of the fast rotators is
dominated by dark matter and of the slow rotators by baryonic matter
(gas and stars). The total mass density (baryonic+dark) is approximately an
order of magnitude higher in the fast rotators. 

5. The implicit assumption $h \sim R$ (Section~\ref{TFdisc-sec}) in
the T-F relation is invalid for dwarf galaxies.  If the velocity at
2.2 optical scale lengths is considered, as is appropriate for larger
spiral galaxies, the deviation of the dwarfs disappears almost
completely.  A comparison with published results for
low-surface-brightness galaxies shows that the larger amplitude of the
halo rotation curve, relative to the maximum allowed stellar rotation
curve is not the reason of this deviation.

6. Three scenarios for evolution in the luminosity-linewidth diagram
were considered : (1). no evolution, (2). evolution in luminosity only,
(3). evolution in both luminosity and linewidth. Evolution to the
T-F relation in luminosity only is discarded because this requires
unrealistically small stellar mass-to-light ratios and because it is
inconsistent with the distribution of baryonic mass in the  
luminosity-linewidth diagram. The deviating dwarf galaxies are either
objects which will never be on the T-F relation (no evolution)
or evolution in both linewidth and luminosity must occur.

\acknowledgements{
This research has made use of the NASA/IPAC Extragalactic Database (NED)   
which is operated by the Jet Propulsion Laboratory, California Institute   
of Technology, under contract with the National Aeronautics and Space      
Administration.
}

\end{document}